\documentclass[a4paper,11 pt]{article}

\usepackage{mathrsfs} \usepackage{graphicx,color} \usepackage{bbm} \usepackage{multirow}

\usepackage{amsmath} %[intlimits]
\usepackage{amssymb}
\usepackage{amsfonts}
\usepackage{latexsym}
\usepackage{amsthm}
\usepackage{varioref}
\usepackage{listings}
\usepackage{float}
\usepackage{rotate}
\usepackage{epsfig}
\usepackage{psfrag}
\usepackage{mathtools}
\usepackage{longtable}

\usepackage{amsmath} 
\usepackage{amsthm}
\usepackage{amssymb}
\usepackage{amsopn}

\begin{document}

\title{ECONOPHYSICS: \\ STILL FRINGE AFTER 30 YEARS?}
\author{JP Bouchaud \footnote{Capital Fund Management, 23 rue de l'Universit\'e, 75007 Paris, France}}

\maketitle

It is customary to date the initial foray of physicists into economics to the September 1987 Santa Fe conference on ``Economics as a complex evolving system", organised by two famous physicists (Phil Anderson \& David Pines) and one famous economist (Ken Arrow).\footnote{There were of course several precursors (for example W. Weidlich).}  When confronted with the theories of Rational Expectations and Efficient Markets, then at the peak of their glory, Phil Anderson famously quipped: {\it Do you guys really believe that?}  Many inspiring and insightful ideas were discussed during that conference [1], suggesting a potentially productive cross-fertilisation between economics and the (back then) nascent theory of complex systems. After all, economics is primarily about inferring the coordination properties of large assemblies of heterogeneous, interacting individuals --- a difficult endeavour that statistical physics has been grappling with for more than a century.

Unfortunately, this cross-fertilisation did not really follow through, at least not on a large scale, in spite of the involvement of larger and larger cohorts of physicists in the development of what is now called ``econophysics" [2] and ``sociophysics" [3]. Until recently, mainstream economists have essentially muddled through along the same well-trodden path, developing monetary policy tools based on the idea that agents are rational calculators with infinite foresight, who optimize their ``utility function" assuming others will do the same. In 2003, Robert Lucas --- one of the dons of Rational Expectations and Nobel Prize recipient in 1995 --- claimed that {\it the central problem of depression prevention ha[d] been solved}. We know all too well what happened in 2008. Jean-Claude Trichet, then Chairman of the European Central Bank, declared: {\it Models failed to predict the crisis and seemed incapable of explaining what was happening [...] in the face of the crisis, we felt abandoned by conventional tools.} Queen Elisabeth II candidly asked why nobody had seen it coming.\footnote{see the answer of UK economists to the Queen: www.feed-charity.org/user/image/besley-hennessy2009a.pdf} Robert Lucas doubled down in 2009 when he insisted that the 2008 crisis was not predicted because economic theory predicts that such events cannot be predicted.

Let's face it: we humans are lost in the dark, using acceptable --- but not optimal --- rules of thumb to cope with complex problems. We strongly interact with one another, either directly or indirectly through market prices or other aggregate indicators while subject to an impressive list of behavioral biases [4] that lead to systematic errors. Analogously to many complex systems they have learnt to deal with [5, 6], physicists believe that all these ingredients should lead to collective effects, instabilities and crises, which may be relevant to economics and finance. It is, I believe, a fair description of the agenda that early econophysicists had in mind, which, owing to the failings of established dogma, seemed like a worthwhile endeavor indeed. This belief is bolstered by a set of empirical results that do not fit well in the framework of classical economics while sounding rather familiar to statistical physicists.

Classical economics is fraught with so-called ``anomalies" and ``puzzles". One of the most famous is the {\it excess volatility puzzle}: the fluctuations of financial markets and of large economies are far too big to be explained by fundamentals alone. In theory, large economic systems should average-out idiosyncratic shocks (say as $N^{-1/2}$ where $N$ is the size of the economy), however, aggregate ``volatility" --- defined as the standard deviation --- is in fact very high. (For example the month-on-month volatility of the US industrial production index since 1950 is as high as $0.75 \%$ around the average growth trend of $0.25 \%$). This of course reflects the propensity of the economy to cycle through booms and busts. But what shocks are responsible for such economic fluctuations? {\it Despite at least two hundred years in which economists have observed fluctuations in economic activity, we still are not sure}, wrote economist John Cochrane in 1994. In fact, the basic mechanism that triggered the 2008 crisis is still hotly debated. 

The standard macroeconomic model (the so-called Dynamic Stochastic General Equilibrium --- DSGE --- model) describes mild fluctuations around a stable equilibrium, and is therefore not well-equipped to deal with endogenous crises (this explains the dismay of J.-C. Trichet, quoted above).
In financial markets, stock prices move by $\sim 2\%$ a day or more, which is much too high to be justified by objective changes of the firms' fundamental value. In fact, only a small fraction of the volatility can be attributed to news; most of the price jumps seem to come out of nowhere. 

What is even more interesting is that the statistics of price changes appears to be universal across many different asset classes (stocks, commodities, currencies, but also derivatives such as options and CDSs, see Fig. 1). The distribution of price changes falls off as a power law, as in many physical systems close to their critical point. The dynamics of markets is strongly intermittent and resembles multifractal turbulent flows. This was first pointed out by Benoit Mandelbrot in the 60s and 70s, and was the starting point of the spree of econophysics papers in the mid-90s (with some overlap with contemporaneous work in financial econometrics [7]). 

The strongly non-Gaussian and universal nature of the fluctuations are further hints that financial markets are mostly driven by self-referential, endogenous effects, and very little by exogenous shocks affecting the fundamental value of firms.

\begin{figure} 
\begin{center}
\includegraphics[scale=0.6]{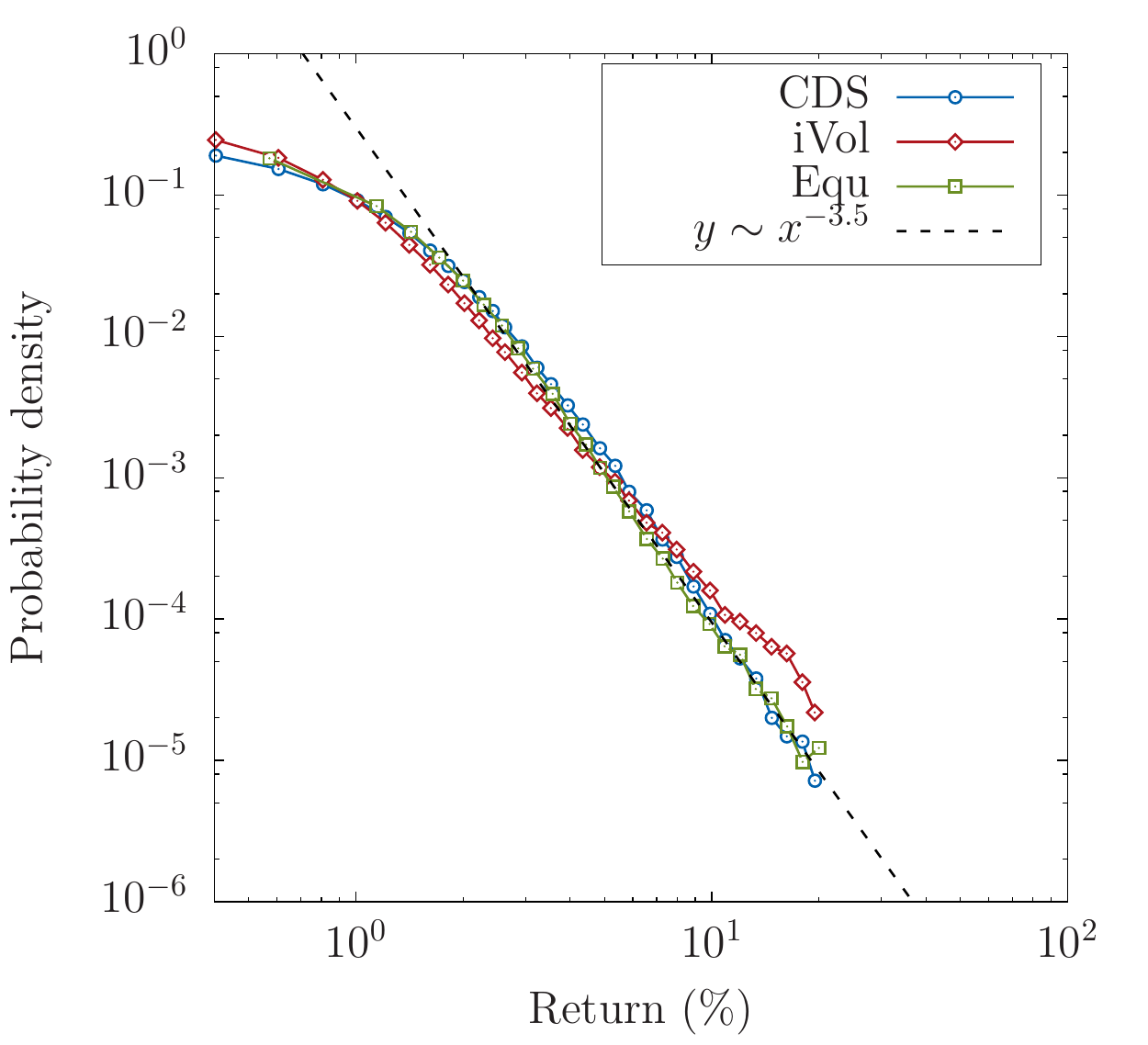}
\caption{Empirical distribution of normalised daily returns of (squares) a
family of US stocks, (circle) the spread of the ``credit default swaps'' (CDS) on the
same stocks, and (diamonds) the so-called ``implied volatilities'' (iVol, related to option markets), again on
the same stocks; all during the period 2011--2013. Returns have been normalised
by their own volatility before aggregation. The tails of these distributions follow
a power-law, with an exponent approximately equal to 3.5.}
\label{fig_pnls}
\end{center}
\end{figure}

Many other fat-tailed distributions have been identified in economics, such as the famous Pareto distribution of wealth and income, or the distribution of firm sizes [8]. For example, a beautiful study by Rob Axtell in 2001 [9] shows that the frequency of firms with a certain number of employees  $S$ decays as $S^{-2}$ over 4 decades. This is highly relevant from a modelling point of view, since the assumption of a ``representative agent/firm" that replaces a whole population by a single average agent or firm might be totally unjustified if the said population is so strongly heterogeneous. 

We know in physics that strong disorder can completely change the nature of the problem (think of Anderson's localisation, or of the fragility of materials when micro-cracks are present). Furthermore, these observations lead to more puzzles for classical economic theories: how can wealth be so unequally distributed when competition is supposed to lead to efficient markets and fair transactions? How can firms be so broadly distributed around a putative equilibrium optimal size?

All these anomalies made for a perfect cocktail in piquing the interest of physicists. Perhaps these reflect interesting (and often universal) underlying collective effects, which are the fundamental mechanism leading to crises and crashes? Quite amazingly, Keynes himself had anticipated such a change of paradigm when he wrote that {\it we are faced at every turn with the problems of Organic Unity, of Discreteness, of Discontinuity; the whole is not equal to the sum of the parts, [...], small changes produce large effects, the assumptions of a uniform and homogeneous continuum are not satisfied.}
 
My own favourite definition of a complex system is precisely that small changes may, with some probability, produce disproportionally large effects, i.e. a system that responds to perturbations in a highly intermittent, non-linear way. The holy grail of econophysics research is to find models that are simple enough to be understood in detail, yet rich enough to generate such a non-trivial phenomenology, with truly unexpected collective phenomena emerging at the macro level.

This may in the end be the most relevant contribution of physics to economics: crises and discontinuities must be understood as interaction-induced, aggregate phenomena that cannot take place in a world of non-interacting, perfectly rational agents.

This agenda is ambitious enough. But what did econophysics actually {\it achieve} in the last 30 years? In terms of total number of papers, quite a lot. In terms of number of very good papers, maybe not that much. In terms of number of very bad papers, certainly too much. In terms of impact on mainstream economics, very little. I don't have the luxury here of reviewing all the work that I particularly value, but, instead, will offer a possible broad classification.

Econophysics has engaged in three types of research projects: 
\begin{enumerate}
\item Empirical studies (for example, as mentioned above, identifying distributions of economic/financial variables); 
\item Technical papers (for example, Random Matrix Theory applied to covariance matrices or Network Theory applied to default contagion) and; 
\item Modelling. The final item can be further decomposed into three categories: phenomenological models, metaphorical models, and realistic (Agent Based) models.\footnote{Of course, the line between these three types of models is blurred, and sometimes phenomenological/metaphorical models are just the first stone needed to build more realistic models.} 
\end{enumerate}

A phenomenological model is mostly descriptive: it attempts to capture in concise mathematical terms the essence of an observation, without necessarily understanding its deeper origin. The Brownian motion as a description of the dynamics of stock markets is the prime example of such models.\footnote{This idea dates back to Bachelier in 1900 and is the most widely used model for prices in financial mathematics. Unfortunately, such a model fails completely at describing the intermittent, fat-tailed nature of stock prices that we described above, see Fig. 1.}

A metaphorical model does not attempt to precisely describe reality, nor does it necessarily rely on very plausible assumptions. Rather, it aims to illustrate a non-trivial mechanism, the scope of which goes much beyond the specifics of the model itself [10]. Well known examples are the Ising model in physics, where the phenomenon of phase transition and collective effects appears in a bare bone model of magnetism, or the Schelling segregation model, that explains how populations can end up living in completely segregated neighbourhoods even though each person individually would not mind living in a mixed neighbourhood. 

Finally, realistic (or Agent-Based) models attempt to build a faithful representation of the world from the bottom up [11] --- like when we derive the laws of thermodynamics and the Navier-Stokes equation from an atomistic description of gases.

Econophysics models fall in the three categories above. Some metaphorical models have, I believe, performed extremely well and are quite influential. One particular example is the Minority Game [12], which is one of the most comprehensive and compelling efforts to show how heterogeneous, competing agents who learn to play a complex game, can generate an extremely rich aggregate behaviour and some of the phenomenology of financial markets. The Schelling metaphorical model can help understand how Adam Smith's invisible hand (another time-worn metaphor) can badly fail when individual agents only follow their self-interest. Simple metaphorical network models for trust evaporation have been devised to help understand how the 2008 crisis unfolded so suddenly, and so violently [13].

Nevertheless, it is a safe bet to think that Agent-Based Models (ABM) will make for the natural common ground for economists and physicists to meet and make progress.\footnote{There is already a well connected community of economists, computer scientists and physicists working on ABMs.}  Starting from realistic models of agents' and firms' behaviour, one is able to generate plausible economies at the macro-scale, with for example spontaneous business cycles and crises without news [14, 15]. These models are also particularly well-suited for conducting monetary policy experiments. For these reasons, ABMs are becoming increasingly popular in both some economics departments and in public institutions (such as the Bank of England [16] and the OECD). 

As models become more realistic and hone in on details, analytics often has to give way to numerical simulations. This is now common practice in physics, where numerical experimentation has gained a respectable status, bestowing on us {\it telescope of the mind,} (as beautifully coined by Mark Buchanan) {\it multiplying human powers of analysis and insight just as a telescope does our powers of vision}. However, the uptake and acceptance of computer simulations as a proper way to do science has been much slower in economics.

Although the economic cost has been dire, the 2008 crisis may have been a blessing for getting theoretical economics out of the rut it had dug for itself. Dissenting voices and alternative, fringe ideas are becoming noticed, even by mainstream economists. 

Calls for ``rebuilding macroeconomics" [17] and rethinking economics undergraduate curricula are now plentiful. For many years, economists' inflexible insistence on Rational Expectations and Efficient Markets made it difficult for physicists to engage in a constructive conversation. This is no longer the case and a window of opportunity is opening up [7], which should not be missed nor botched. 

Real cross-disciplinary efforts have actually started and will hopefully proliferate in years to come --- which means that universities should actively foster ``double curricula" that mix (statistical) physics and economics at an early stage. This is the only way econophysics will not remain on the fringe for the next thirty years.

\newpage
\textbf{References}
\vskip 0.5cm

[1] See the conference proceedings: P. W. Anderson, K. Arrow, D. Pines, The Economy as an Evolving Complex System, Addison-Wesley (1988)

[2] Mantegna, R. N., \& Stanley, H. E. (1999). Introduction to econophysics: correlations and complexity in finance. Cambridge University Press.

[3] see e.g. Castellano, C., Fortunato, S., \& Loreto, V. (2009). Statistical physics of social dynamics. Reviews of modern physics, 81(2), 591.

[4] see: https://en.wikipedia.org/wiki/List-of-cognitive-biases

[5] P. Bak, How Nature Works: The Science of Self-Organized Criticality, New York: Copernicus (1996)

[6] J. Sethna, K. Dahmen, C. Myers, Crackling Noise, Nature, 410, 242 (2001)

[7] on this point, see the very interesting discussion in Jovanovic, F., \& Schinckus, C. (2017). Econophysics and Financial Economics: An Emerging Dialogue. OUP.

[8] for a review: Gabaix, X. (2009). Power laws in economics and finance. Annu. Rev. Econ., 1(1), 255-294.

[9] R. Axtell, (2001) Zipf distribution of US firm sizes. Science, 293(5536), 1818-1820

[10] for a partial review of such models, see: Bouchaud, J. P. (2013). Crises and collective socio-economic phenomena: simple models and challenges. J. Statistical Physics, 151(3-4), 567-606, and Sornette, D. (2014). Physics and financial economics (1776---2014): puzzles, Ising and agent-based models. Reports on progress in physics, 77(6), 062001.

[11] for very insightful introductions to ABMs, see Epstein, J. M. (1999). Agent-based computational models and generative social science. Complexity, 4(5), 41-60; Goldstone, R. L., \& Janssen, M. A. (2005). Computational models of collective behavior. Trends in cognitive sciences, 9(9), 424-430.

[12] D. Challet, M. Marsili, Y.C. Zhang, Minority Games, Oxford University Press (2005)

[13] see e.g. Anand, K., Gai, P., \& Marsili, M. (2012). Rollover risk, network structure and systemic financial crises. J. Economic Dynamics and Control, 36(8), 1088-1100; da Gama Batista, J., Bouchaud, J. P., \& Challet, D. (2015). Sudden trust collapse in networked societies. European Physical Journal B, 88(3), 55.

[14] Delli Gatti, D., Palestrini, A., Gaffeo, E., Giulioni, G., \& Gallegati, M. (2008). Emergent macroeconomics: an agent-based approach to business fluctuations. Springer.

[15] Gualdi, S., Tarzia, M., Zamponi, F., \& Bouchaud, J. P. (2015). Tipping points in macroeconomic agent-based models. J. Economic Dynamics and Control, 50, 29-61.

[16] Haldane, A. G., \& Turrell, A. E. (2018). An interdisciplinary model for macroeconomics. Oxford Review of Economic Policy, 34(1-2), 219-251.

[17] ``Rebuilding Macroeconomics", Oxford Review of Economic Policy, 34(1-2).

\end{document}